\documentclass[preprints,article,accept,moreauthors,pdftex]{Definitions/mdpi}

\firstpage{1} 
\pubvolume{xx}
\issuenum{1}
\articlenumber{5}
\pubyear{2019}
\copyrightyear{2019}
\history{}

\pdfoutput=1

\Title{Medidas de distanciamento social e mobilidade na América do Sul durante a pandemia por COVID-19: Condições necessárias e suficientes?}

\Author{Gisliany Lillian Alves de Oliveira$^{1}$\orcidG{}, Luciana Conceiç\~{a}o de Lima$^{2}$\orcidL{}, Ivanovitch Silva$^{3}$\orcidI{}, Marcel da Câmara Ribeiro-Dantas$^{4}$\orcidM{}, Kayo Henrique Monteiro$^{5}$\orcidK{}, Patricia Takako Endo $^{6}$\orcidP{}}

\address{%
$^{1}$ \quad Universidade Federal do Rio Grande do Norte; gisliany.alves@ufrn.edu.br\\
$^{2}$ \quad Universidade Federal do Rio Grande do Norte; luciana.lima@ccet.ufrn.br\\
$^{3}$ \quad Universidade Federal do Rio Grande do Norte; ivan@imd.ufrn.br\\
$^{4}$ \quad Institut Curie; marcel.ribeiro-dantas@curie.fr\\
$^{5}$ \quad Universidade de Pernambuco; khcm@ecomp.poli.upe\\
$^{6}$ \quad Universidade de Pernambuco; patricia.endo@upe.br}

\abstract{Em um cenário de inexistência de vacina para a COVID-19, intervenções não farmacêuticas são necessárias para conter o espraiamento do vírus e o colapso do sistema de saúde nas regiões afetadas. Uma dessas medidas é o distanciamento social, que objetiva reduzir as interações na comunidade por meio de fechamentos de estabelecimentos públicos e privados que envolvam aglomerações de pessoas. O \textit{lockdown} pressupõe a redução drástica das interações comunitárias, representando uma medida mais extrema de distanciamento social. Com base em dados de geolocalização disponibilizados pela Google para seis categorias de espaços físicos, este artigo identifica as variações na circulação de pessoas na América do Sul para diferentes modalidades de distanciamento social adotadas durante a pandemia por COVID-19. Para este trabalho, foram analisadas as tendências de mobilidade da população para um conjunto de países entre 15 de fevereiro de 2020 e 16 de maio de 2020. Para sumarizar essas tendências em uma única métrica, criou-se um indicador geral de circulação, e para identificar padrões regionais de mobilidade foram utilizadas análises descritivas de autocorrelação espacial (índice de Moran global e local). A primeira hipótese deste trabalho é de que países com decreto de \textit{lockdown} podem obter maior sucesso em reduzir a mobilidade da população, e a segunda hipótese é de que Argentina, Brasil e Colômbia apresentam padrões regionais de mobilidade. A primeira hipótese foi parcialmente confirmada (considerando 10 países da América do Sul), e os resultados obtidos nas análises espaciais confirmaram a segunda hipótese. No geral, nos dados observados verifica-se que medidas de \textit{lockdown} ou de distanciamento social menos rígidas são necessárias, porém, não são suficientes para se alcançar uma redução significativa da circulação de pessoas durante a pandemia. 
}


\begin{document}


\section{Introdução}
No final de 2019, autoridades chinesas começaram a reportar casos de pneumonia por etiologia desconhecida na cidade de Wuhan, localizada na província de Hubei, região central daquele país. Dentro de poucas semanas, outras nações da região notificaram casos similares, indicando que um novo tipo de coronavírus se encontrava em circulação, o o SARS-CoV-2. No dia 11 de março de 2020, a Organização Mundial da Saúde (OMS) declarou pandemia global pela COVID-19 \footnote{\url{https://www.who.int/dg/speeches/detail/who-director-general-s-opening-remarks-at-the-media-briefing-on-covid-19---11-march-2020}}. Enquanto nas primeiras semanas de março de 2020, a China apresentava tendência de estabilização do número de casos de pessoas infectadas pelo SARS-CoV-2~\cite{italia1, italia2}, a Itália se destacava como epicentro mundial da doença~\cite{italia2}. Diante de um cenário crítico de óbitos e casos confirmados pelo novo coronavírus, o governo italiano endureceu seus decretos de distanciamento social, o que parece ter produzido efeitos sobre o ritmo de avanço da COVID-19 que tem apresentado sinais de redução desde abril de 2020 \cite{de2020covid}. Outros países europeus como Espanha, Alemanha, França e Suécia também foram duramente atingidos pela pandemia por COVID-19 \cite{singh2020prediction}.

Nas Américas, até o dia 6 de junho de 2020 as cinco nações com os maiores número de casos confirmados eram, por ordem, os Estados Unidos, o Brasil, o  Peru, o Chile e o México \cite{WHO-136}. Também com relação a essa data, o número de novos casos e óbitos no continente americano eram expressivos, perdendo apenas para a Europa \cite{WHO-136}.

Considerando as especificidades socioeconômicas da América do Sul e os ritmos distintos de avanço da doença dentro da região, o presente artigo destaca um aspecto central para o sucesso das medidas que visam conter a propagação do vírus: o distanciamento social. De acordo com os especialistas, ações como o suspensão das atividades escolares,  prática de \textit{home office}, redução no uso de transporte público, cancelamento de eventos esportivos e demais atividades que envolvam um grande número de pessoas, e a utilização de tecnologias digitais para a comunicação diária, podem fazer a diferença no nível de doentes e mortos pela doença \cite{lippi2020health, the2020covid}. Quanto menor a interação social entre as pessoas, mais distribuído será o número de novos casos ao longo do tempo, evitando dessa forma que o sistema de saúde entre em colapso em um cenário de excesso de demanda por atendimentos médicos para o tratamento da COVID-19 \cite{lippi2020health, the2020covid, sjodin2020only}. 

O \textit{lockdown} é o caso mais extremo de distanciamento social, cujo objetivo é reduzir a interação entre as pessoas de maneira mais contundente, por meio de normativas que passam pela restrição na circulação dentro de uma determinada localidade, seja ela total ou parcial \cite{pescarinimedidas, srivastava2020global}. Há estudos que se voltam para a análise do impacto das medidas de distanciamento sobre a circulação de pessoas em diversas regiões do mundo, sem contudo realçar a América do Sul \cite{lasry2020timing}. No entanto, para essa região, que no mês de maio passou a representar o epicentro da COVID-19, estudos nessa linha ainda são escassos e precisam avançar. 
Tendo em vista o perfil diverso da região da América do Sul quanto às medidas adotadas para enfrentamento do novo coronavírus, assim como no que se refere aos aspectos culturais, socioeconômicos e políticos, o presente estudo estabelece duas hipóteses: a) países que adotaram o \textit{lockdown} desde o início da pandemia são mais efetivos em reduzir a mobilidade da população; b) nos países selecionados da América do Sul há a formação de padrões regionais de mobilidade. 

Conhecer o impacto das medidas de distanciamento social na América do Sul, e cujas medidas mais extremas têm sido a regra, pode auxiliar na compreensão da eficiência de intervenções não farmacêuticas sobre o controle da circulação de pessoas, que por sua vez, afeta o nível de transmissão da doença na comunidade.

Além dessa introdução, o presente trabalho se subdivide em mais três seções: em Materiais e Métodos são apresentadas as fontes de dados e as técnicas de análise utilizadas para a consecução dos objetivos propostos. Em seguida, são apresentados os resultados obtidos para um conjunto de dez países da América do Sul, enquanto a discussão sobre as principais descobertas e pontos evocados são descritos na última seção.

\section{Materiais e Métodos}
\subsection{Dados do Google Community Mobility Report}
Desde abril de 2020, a Google tem disponibilizado em seu Google Community Mobility Reports~\footnote{\url{https://www.google.com/covid19/mobility/}} informações sobre locais visitados e capturados por dispositivos móveis para 132 países. Para que os locais visitados sejam registrados, o usuário precisa ativar o Histórico de Localização de sua conta, que por padrão se encontra desativado. As informações são disponibilizadas como séries históricas em arquivos de formato .csv (registros anônimos e agregados) para cada país e possui as seguintes categorias de espaços físicos: (a)compras e recreação (shoppings centers, restaurantes, livrarias); (b)supermercados e farmácias; (c)áreas de lazer (parques, praias, jardins e praças públicas); (d)estações de trânsito (pontos de embarque e desembarque de passageiros); (e)locais de trabalho; e residências dos usuários. A Figura \ref{fig:dados-google} ilustra a estrutura da base de dados de mobilidade disponibilizado pela Google:

\begin{figure}[H]
    \centering
    \includegraphics[width=1.0\textwidth]{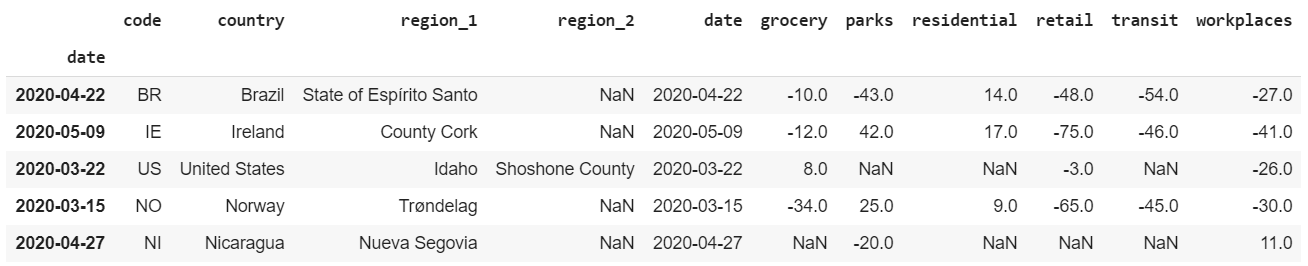}
    \caption{Base de dados de relatórios de mobilidade da comunidade da Google. Fonte dos dados básicos: COVID-19 Community Mobility Reports, Google.}
    \label{fig:dados-google}
\end{figure}

Os dados divulgados pela Google são percentuais de variação na mobilidade da população, tendo como linha de base a movimentação no período de 3 de janeiro a 6 de fevereiro de 2020 (período considerado pré-pandemia na região). 

\subsection{Seleção dos países da América do Sul}
Para a análise das tendências de mobilidade dos países da América do Sul, foram selecionadas nações para as quais se tivessem dados para o período de 15 de fevereiro a 16 de maio. Com base nesse critério incluiu-se 10 países sul-americanos: Argentina, Bolívia, Brasil, Chile, Colômbia, Equador, Paraguai, Peru, Uruguai e Venezuela. 

Desse conjunto foram selecionadas três nações para proceder às análises de autocorrelação espacial: Argentina, Brasil e Colômbia. Elas foram escolhidas por formarem um grupo misto quanto à adoção de medidas de distanciamento social mais e menos rígidas. Argentina e Colômbia adotaram medidas de \textit{lockdown} no início da pandemia (20 e 24 de março, respectivamente), enquanto o Brasil seguia um  decreto nacional de distanciamento social menos rígido, apesar de alguns municípios brasileiros apresentarem decreto de \textit{lockdown} ao final do período analisado, mais precisamente a partir de maio. Ademais, para essas três nações as informações de mobilidade também se encontravam desagregadas para unidades menores de análise, no caso, províncias/departamentos e estados (do total de 132 países da base de dados da Google essas características estavam presentes em apenas 39\% deles). Essa foi uma condição necessária para que se pudesse identificar agrupamentos de mobilidade no interior dos países selecionados.

Para a utilização de informações de províncias/departamentos e estados foi preciso lidar com as informações faltantes de registros de mobilidade por categorias. No caso da Argentina, elas ocorreram apenas para a categoria representativa das estações de trânsito e a uma taxa de 1,49\%. Para a Colômbia, diferentes percentuais de dados ausentes foram encontrados em praticamente todas as categorias: locais de trabalho (2,07\%), compras e recreação (7,35\%), supermercados e farmácias (7,35\%), residências (18,28\%) e estações de trânsito (20,60\%). 

A fim de tratar esses dados omissos, optou-se por um método de imputação única pela média dos valores por país e por categoria, escolha que se justifica na simplicidade de implementação do método para gerar o conjunto de dados completo \cite{shryock1973methods}.

\subsection{Indicador de circulação}
\begin{figure}[htpb]
    \centering
    \includegraphics[width=0.7\textwidth]{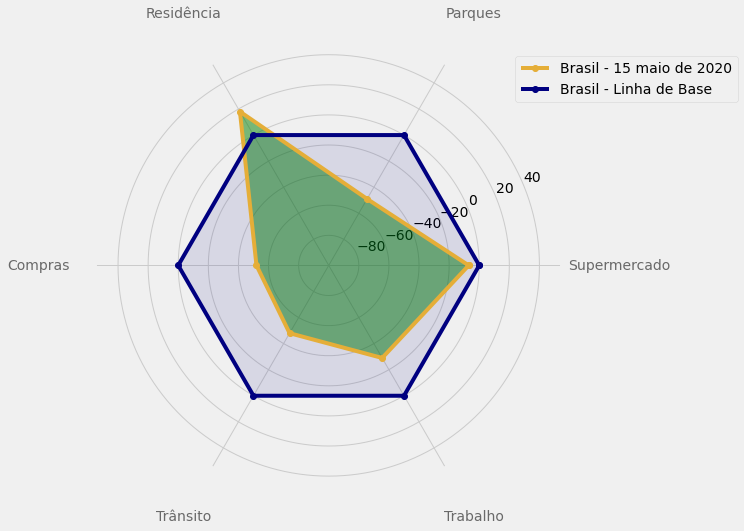}
    \caption{Cálculo do indicador de circulação de área sobre um gráfico de radar. Fonte dos dados básicos: COVID-19 Community Mobility Reports, Google.}
    \label{fig:indicador-radar}
\end{figure}

Para sumarizar as tendências de mobilidade das seis categorias de atividades da Google em uma única métrica, criou-se um indicador de circulação com base na área de um gráfico de radar\cite{limappgdem}. Conforme pode-se observar na Figura \ref{fig:indicador-radar}, cada triângulo que compõe o hexágono forma uma área que pode ser calculada pela lei dos senos e que tem como ponto central um valor mínimo, $C$ \cite{limappgdem}. O indicador de circulação é então obtido pela razão entre o somatório das áreas dos seis triângulos no período analisado e o somatório das áreas dos seis triângulos no período de 3 de janeiro a 6 de fevereiro de 2020 (linha de base), anterior à chegada da pandemia por COVID-19 na América do Sul. 

Tendo como referência à linha de base, valores do indicador menores que um, indicam menor nível de circulação de pessoas; iguais a um, mesmo nível de circulação de pessoas; e maiores que um, maior nível de circulação de pessoas. 

A fim de se retirar os efeitos dos dias das semana, as tendências foram dessazonalizadas por meio da aplicação de técnicas estatísticas (\textit{Season-Trend decomposition using LOESS})~\footnote{Para mais detalhes, ver: \url{https://www.statsmodels.org/devel/generated/statsmodels.tsa.seasonal.STL.html}}.

\subsection{Análise de autocorrelação espacial}
Uma das hipóteses do presente estudo é que a adesão ao distanciamento social não se distribui de maneira aleatória no espaço. Portanto, é possível que durante a pandemia por COVID-19 padrões regionais de mobilidade na América do Sul tenham se formado.

Cada variação de mobilidade do conjunto de dados está associada a recortes espaciais em duas diferentes granularidades: nacional e suas subdivisões político-administrativas (estados ou províncias/departamentos). Sendo assim, para avaliar a hipótese mencionada, a análise da autocorrelação espacial é capaz de revelar se existe alguma associação entre essas variações de mobilidade próximas no espaço, se elas formam agrumentos ou se ocorrem ao acaso.
A autocorrelação espacial pode ser entendida como uma avaliação correlacional dos atributos de uma mesma variável em diferentes pontos no espaço, a fim de verificar como as magnitudes desses valores influenciam e são influenciados pelos locais vizinhos \cite{anselin2009approaches}. Quando a autocorrelação é positiva ocorre similaridade espacial, isto é, valores próximos no espaço também se assemelham em seus valores, formando agrupamentos (\textit{clusters}). Quando a autocorrelação é negativa, essa relação entre proximidade e valores de uma variável se inverte e há heterogeneidade espacial \cite{dube2014spatial}. Há ainda o caso da aleatoriedade espacial, caracterizada pela inexistência de qualquer relação significativa entre a localização das variáveis e seus atributos \cite{anselin2009approaches}.

\begin{figure}[H]
    \centering
    \includegraphics[width=0.6\textwidth]{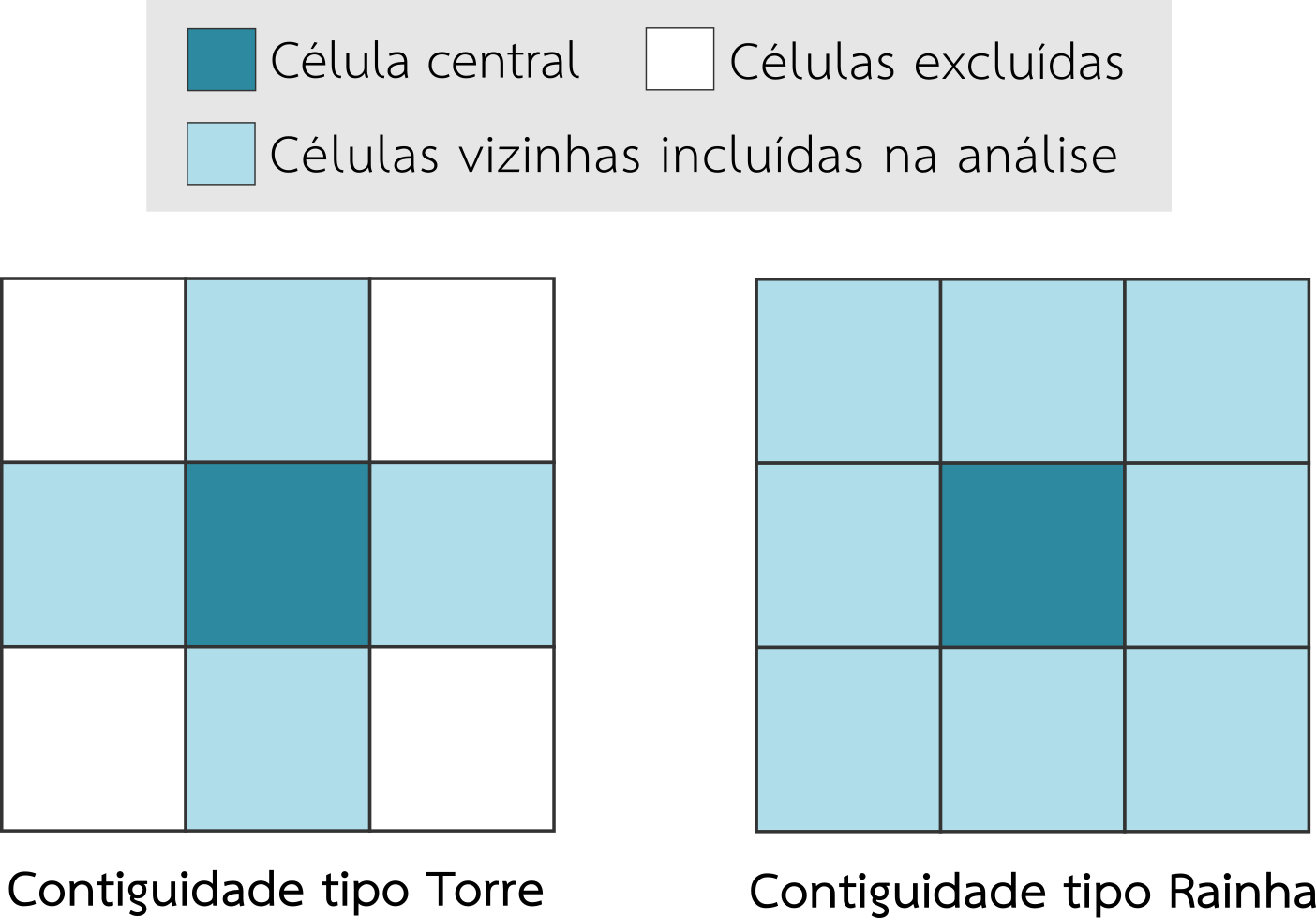}
    \caption{Critérios de vizinhança por contiguidade tipo Torre e Rainha. Fonte: Adaptado de \cite{lloyd2010spatial}.}
    \label{fig:torre-rainha}
\end{figure}

Para estimar a autocorrelação, uma das técnicas estatísticas mais comuns é o Teste de Moran, que apresenta um índice global para avaliar toda a região em análise e índices locais para os recortes distintos dentro de uma mesma região \cite{anselin2009approaches, dube2014spatial, chi2008spatial}. Ambos os índices levam em consideração a influência da vizinhança, representada por meio de uma matriz de proximidade espacial $W$, em que cada elemento $w_{ij}$ mede a proximidade entre as área $i$ e $j$ por algum critério de contiguidade espacial \cite{dube2014spatial, monteiro2004analise}. As especificações de vizinhança mais utilizadas são a contiguidade do tipo Torre, em que se define que áreas vizinhas compartilham apenas bordas, e a do tipo Rainha, em que locais vizinhos compartilham bordas e vértices, e que foi a utilizada no presente estudo \cite{feola2017diffusion}.Esses critérios de vizinhança são ilustrados na Figura \ref{fig:torre-rainha}.

De acordo com \cite{anselin2009approaches}, o índice de Moran global, $I$, para uma variável $x$ pode ser descrito formalmente conforme a Equação~\ref{eq:moran-global}.

\begin{equation}\label{eq:moran-global}
    I = \frac{n}{S_0}\frac{\sum_{i}\sum_{j} w_{ij}(x_i - \mu)(x_j - \mu)}{(x_i - \mu)^2}  
\end{equation}

Onde $n$ é o número de áreas, $S_0$ é a soma de todos os elementos da matriz de proximidade espacial $w$ e $\mu$ é a média de todas as observações de $x$. O índice de Moran, $I$, varia entre -1 e 1, com $I>0$ indicando autocorrelação positiva e $I<0$ apontando para uma autocorrelação negativa, havendo ainda sua validação por meio de um teste de pseudo-significância cuja hipótese nula é a independência espacial ($I=0$)
~\cite{anselin2009approaches}.

\begin{figure}[H]
    \centering
    \includegraphics[width=0.5\textwidth]{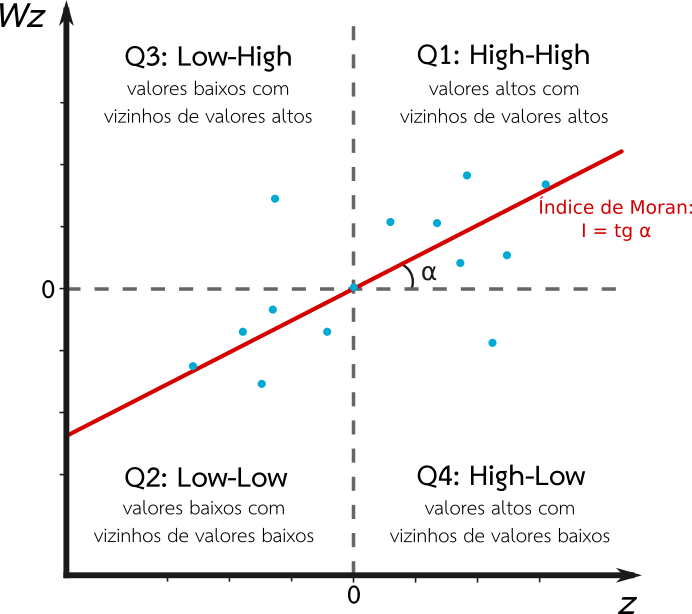}
    \caption{Diagrama de Espalhamento de Moran. Fonte: Adaptado de~\cite{amaral2000,gutoiu2015}.}
    \label{fig:moran-scatterplot}
\end{figure}

Uma interpretação visual do índice de Moran é a inclinação da reta de regressão do chamado Diagrama de Espalhamento de Moran~\cite{anselin2009approaches}, conforme ilustrado na Figura~\ref{fig:moran-scatterplot}. Esse diagrama consiste em um gráfico bidimensional, em que uma variável normalizada $z$ é disposta sobre o eixo x para ser comparada com a média da sua vizinhança $W_z$ (em que $W$ se encontra normalizada, i.e., com suas linhas redimensionadas para somar um), disposta sobre o eixo y~\cite{gutoiu2015}. Logo, os pontos nos quadrantes $Q_1$ e $Q_2$ representam áreas com valores semelhantes (autocorrelação positiva), enquanto os pontos dos quadrantes $Q_3$ e $Q4$ ilustram áreas com valores dissimilares (autocorrelação negativa)~\cite{barreca2017}.

A visualização desses dados por meio de mapas é desejável. Nesse caso, os recortes espaciais são coloridos conforme o quadrante a que cada ponto pertence no diagrama de Moran~\cite{griffith2009spatial}. Como o diagrama de espalhamento de Moran não oferece informação sobre a significância de cada observação dentro de seus quadrantes, torna-se necessário investigar essa estatística por meio do uso de Indicadores Locais de Associação Espacial (\textit{Local Indicatos of Spacial Association} ou LISA)~\cite{barreca2017}. O índice de Moran local é parte da classe de indicadores LISA e traz um detalhamento dos padrões e tendências locais~\cite{griffith2009spatial}, sendo definido para cada área $i$ a partir dos valores da variável $z_i$ conforme a Equação~\ref{eq:moran-local}.

\begin{equation}\label{eq:moran-local}
    I_i = z_i \sum_j w_{ij}z_j
\end{equation}

Em que os elementos $w_{ij}$ da matriz de proximidade espacial estão normalizados e o somatório em $j$ engloba apenas os valores vizinhos $j \in J_i$ da área $i$~\cite{barreca2017}. Como os índice de Moran local também apresentam um teste estatístico para sua validação análogo ao do índice global~\cite{anselin2009approaches}, torna-se viável que a visualização em mapas destaque apenas os recortes espaciais com autocorrelação espacial significativa. Na literatura, esses mapas são conhecidos como mapas de \textit{cluster} LISA e são geralmente acompanhados de outros mapas análogos que destacam apenas o nível de significância de cada recorte, chamados de mapas (de significância) LISA~\cite{barreca2017,fontes2014accessibility}.

\section{Medidas de distanciamento social na América do Sul e circulação de pessoas}
\begin{figure}[H]
    \centering
    \includegraphics[width=0.9\textwidth]{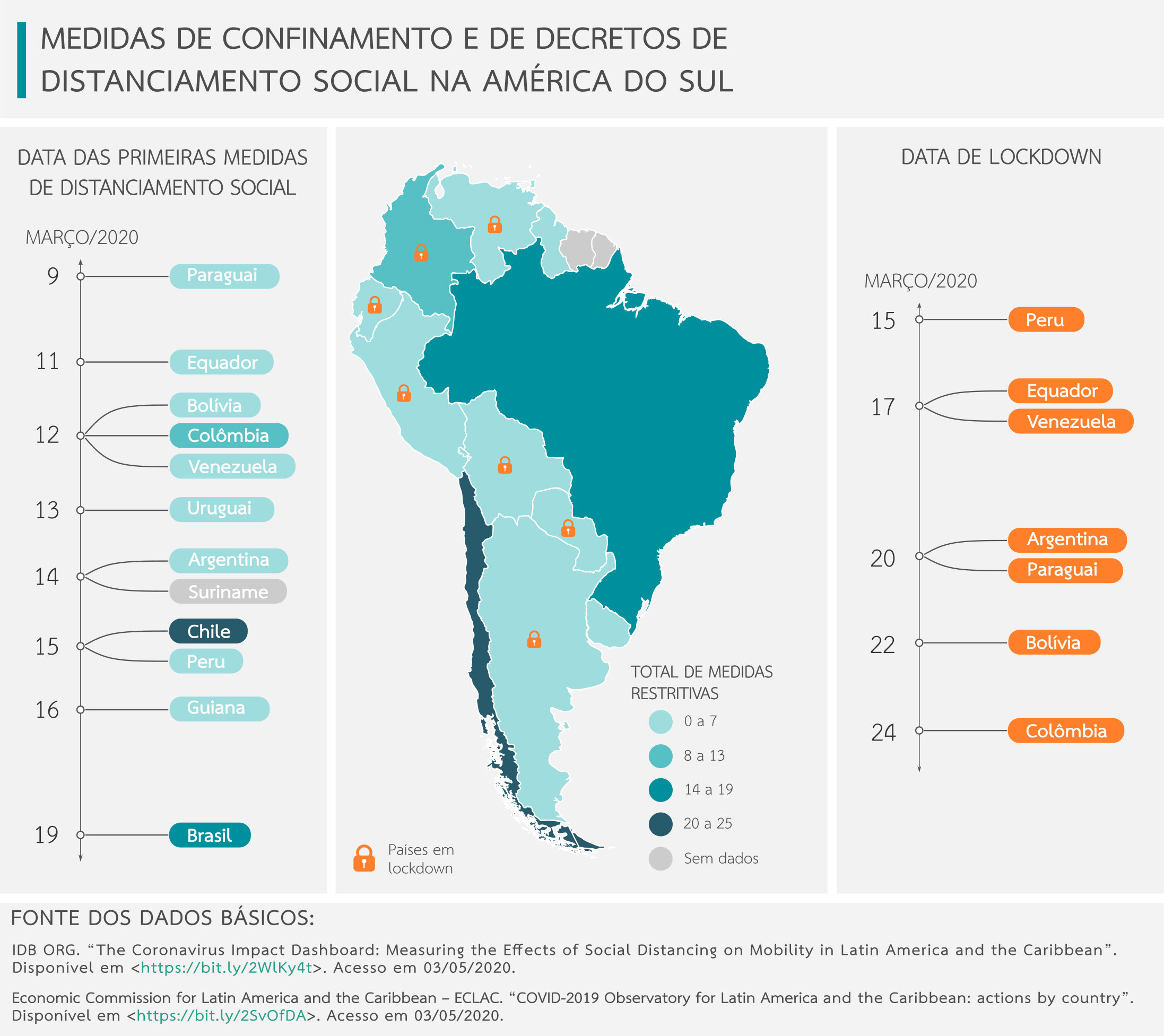}
    \caption{Medidas de \textit{lockdown} e de decretos de distanciamento social na América do Sul considerando a data limite de 25 de março.}
    \label{fig:america-sul-lockdown}
\end{figure}

Conforme apresentado na Figura \ref{fig:america-sul-lockdown}, o Peru foi o primeiro país da América do Sul a adotar o \textit{lockdown} (15 de março), seguido por Equador e Venezuela (17 de março), Argentina e Paraguai (20 de março), Bolívia (22 de março) e Colômbia (24 de março). Até o dia 25 de março, estavam em vigor no Brasil, no Uruguai, no Chile e na Guiana medidas de distanciamento social menos rígidas, embora no caso específico do Chile, medidas de fechamento de algumas áreas de seu território tenham sido adotadas \cite{eclac}. 

Quanto ao número de medidas de distanciamento social aplicadas para contenção do avanço do novo coronavírus, se destacaram o Chile (25 medidas) e o Brasil (14 medidas), dois países que não optaram pelo \textit{lockdown} em todo o seu território até o dia 25 de março (Figura \ref{fig:america-sul-lockdown}). No caso do Chile predominaram medidas nacionais relacionadas às restrições ou fechamentos de locais públicos ou de circulação de grande número de pessoas (13 no total). No Brasil, essa mesma restrição foi tomada e, junto com a restrição ou proibição da entrada de estrangeiros no país, somaram 12 medidas \cite{eclac}.

\section{Medidas de distanciamento social e circulação de pessoas na América do Sul: há um padrão?}
\begin{figure}[ht]
    \centering
    \includegraphics[width=1.0\textwidth]{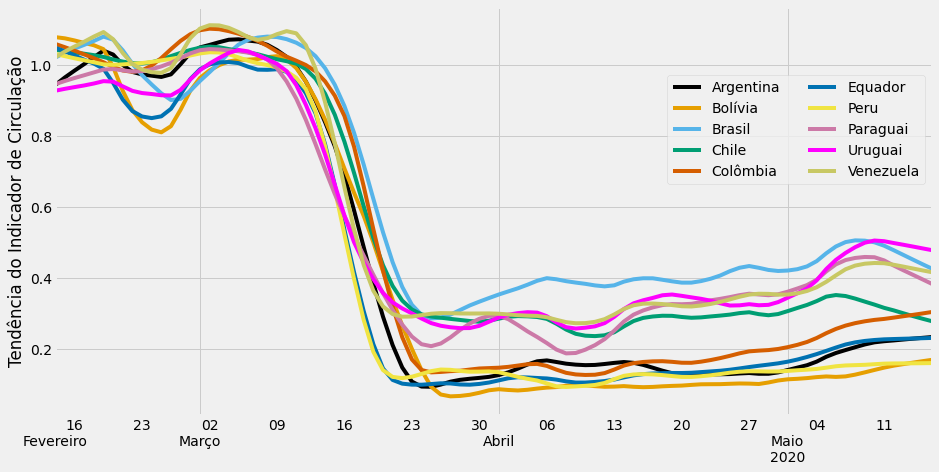}
    \caption{Evolução do indicador de circulação na América do Sul, 15 de fevereiro a 16 de maio.}
    \label{fig:evolucao-circulacao}
\end{figure}

A Figura \ref{fig:evolucao-circulacao} ilustra a trajetória seguida pelo indicador de circulação da Google entre 15 de fevereiro e 16 de maio para o grupo dos 10 países sul-americanos em análise. Observa-se uma queda acentuada da circulação de pessoas capturadas pela Google nas datas coincidentes com as comemorações de carnaval (na segunda quinzena de fevereiro). A tendência de redução na mobilidade começou aproximadamente a partir da segunda semana de março, quando boa parte dos países analisados anunciaram as primeiras medidas de distanciamento social. 

Nota-se uma velocidade maior de declínio de circulação para nações como o Paraguai, primeiro país da América do Sul a decretar medidas de distanciamento social (09 de março), e também para o Uruguai, cujas medidas iniciais foram decretadas em 13 de março. Também chama a atenção o caso da Venezuela:  em 12 de março, esse país aplicou suas primeiras medidas relacionadas à restrição de circulação de pessoas, porém, os níveis de mobilidade permaneceram como um dos mais elevados por mais alguns dias, até alcançar um ritmo de declínio com o decreto de \textit{lockdown} em 17 de março, quatro dias após esse país confirmar o seu primeiro caso de COVID-19 \cite{paniz2020arrival}.

A partir de 23 de março, aproximadamente, nota-se para a maior parte do período analisado a formação de dois grupos: o primeiro, com menor circulação, incluiu nações que declararam \textit{lockdown} ainda no início da pandemia; o segundo, com maior circulação, incluiu países que adotaram as duas modalidades de distanciamento social: 

\subsection{Países com menores níveis de mobilidade da população}
A Bolívia foi o país com o menor nível de mobilidade da população segundo o indicador analisado, embora tenha apresentado uma tendência de crescimento ao final do período. A Colômbia, que vinha apresentando baixos níveis de mobilidade  iniciou ritmo de aumento em sua tendência a partir da segunda quinzena de abril, aproximadamente. Esse país terminou a série analisada como um dos cinco países da América do Sul com o maior nível de circulação de pessoas, o que é compatível com as medidas de relaxamento adotadas em maio~\footnote{\url{https://colombiareports.com/colombia-extends-lockdown-until-may-31-and-health-emergency-until-august-31/}} que flexibilizaram alguns pontos do decreto original de \textit{lockdown}.

No caso da Argentina, que decretou o distanciamento social mais rígido em 20 de março, esse país se situou entre os países com menores níveis de circulação da região, com tendência de crescimento da mobilidade em maio. O que é compatível com medidas de relaxamento de algumas restrições estabelecidas e em determinadas localidades do território argentino. 

Com relação ao Equador, a epidemia avançou fortemente nesse território desde a confirmação do primeiro caso, no final de fevereiro \cite{hallo2020perspective}. Foram decretadas medidas de contenção da circulação de pessoas ainda nas primeiras semanas de março, sobretudo na capital Quito \cite{movilidadquito}. O Equador que possui uma população de cerca de 17 milhões habitantes~\footnote{\url{https://data.worldbank.org/indicator/SP.POP.TOTL}} contabilizava até o dia 6 de junho mais de 3 mil mortes pela COVID-19 \cite{WHO-136}.

\subsection{Países com maiores níveis de mobilidade da população}
Considerando os países que apresentaram maior nível de circulação no período, o Brasil foi o destaque, tendo se mantido na maior parte do período pós-pandemia nos maiores patamares em relação às demais nações analisadas. Aproximadamente a partir de maio observa-se para esse país tendência de redução na circulação de pessoas, o que pode ser uma decorrência das medidas de \textit{lockdown} tomadas de maneira independente por alguns municípios importantes, sobretudo na Região Nordeste. O Brasil entrou no mês de maio contabilizando mais óbitos pela COVID-19 do que a China, primeiro país do mundo a registrar casos pela doença, ainda no final de 2019. Segundo o Ministério da Saúde daquele país, em 01 de maio, o número de mortes pelo novo coronavírus no Brasil era igual a 7.321, e o número de infectados igual a 107.780 \cite{coronavirusbrasil}. E esse aumento do ritmo de propagação da doença fomentou as discussões e adesão ao \textit{lockdown}, e à parte das diretrizes do governo federal sobre o distanciamento social. 

No caso do Chile, chama a atenção a tendência seguida pelo que a partir do mês de maio parece ter alcançado ritmo sustentado de declínio da circulação de pessoas, segundo o indicador analisado. Foi decretado \textit{lockdown} na região metropolitana de Santiago (capital administrativa do Chile) a partir de 15 de maio~\footnote{\url{https://agenciabrasil.ebc.com.br/saude/noticia/2020-05/chile-decreta-lockdown-em-santiago-apos-explosao-de-casos-de-covid-19}}, em decorrência da elevação do número de casos da doença, e isso pode ter impactado a mobilidade no Chile algumas semanas antes da tomada dessa medida extrema de distanciamento social. 

Também entre os países com maior nível de circulação de pessoas, destacou-se o Uruguai. No fim da série analisada esse país superou o Brasil em termos de circulação de pessoas, o que coincide com o período de retomada de setores da economia uruguaia.

No Paraguai, que adotou medidas de \textit{lockdown} em 20 de março, o comportamento de mobilidade da população apresentou variações significativas: por vezes se aproximando do nível de redução de circulação dos países do segundo grupo, mas terminando a série analisada com tendência mais próxima ao dos países com os maiores níveis de circulação e que não adotaram medidas mais rígidas de distanciamento social. Com base nesses resultados, a hipótese de que em países com decreto de \textit{lockdown} há maior sucesso em reduzir a mobilidade da população foi parcialmente confirmada, tendo em vista que ao menos para um país analisado ela não foi corroborada.
 
A próxima subseção apresenta as análises espaciais descritivas para três países selecionados e suas divisões territoriais em províncias/departamentos e estados: Argentina, Colômbia e Brasil.

\section{Análises espaciais das tendências internas de mobilidade}
\subsection{Análise exploratória de dados espaciais}
\begin{figure}[ht!]
    \centering
    \includegraphics[width=0.9\textwidth]{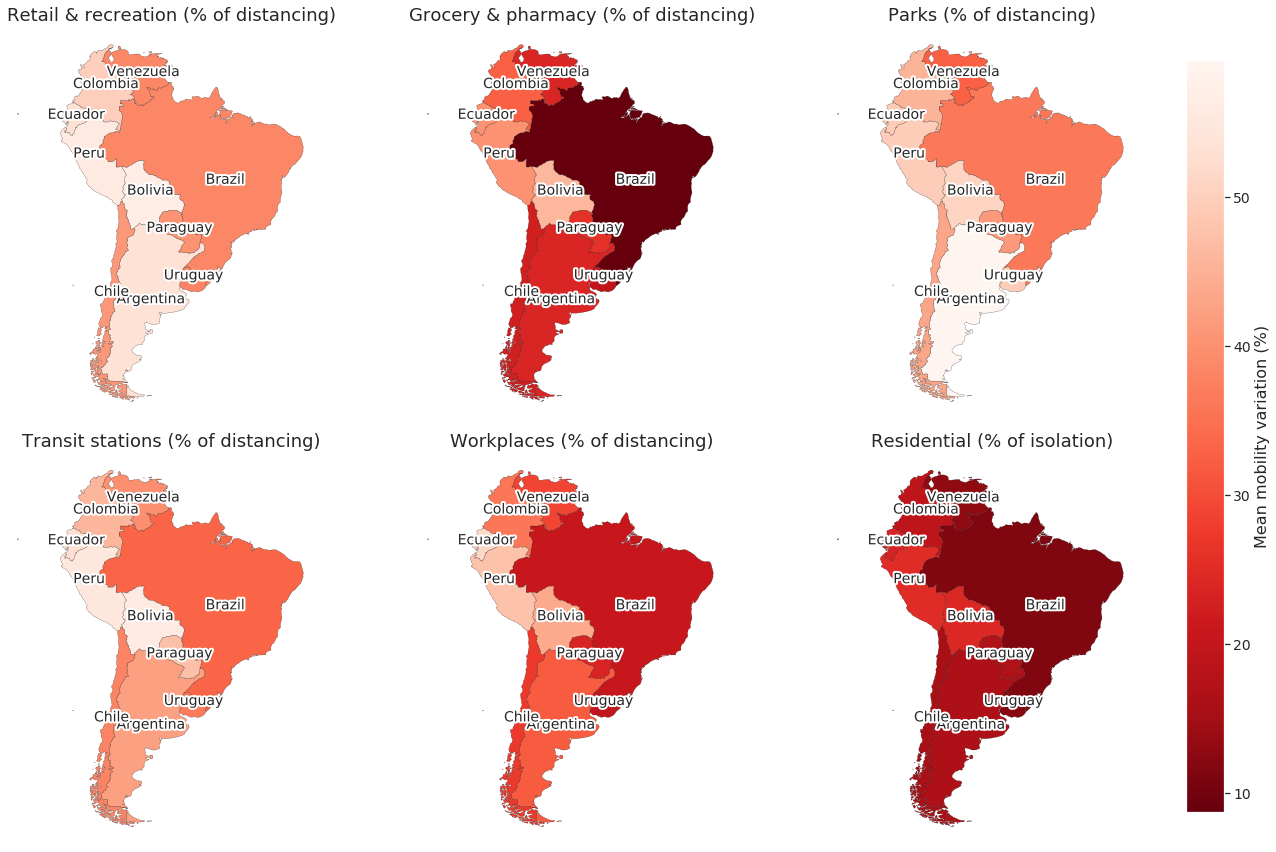}
    \caption{Variação média da mobilidade em relação à linha de base (3 de janeiro a 6 de fevereiro de 2020) para países selecionados da América do Sul durante o período de 15 de fevereiro a 16 de maio de 2020. Cores mais claras indicam maior distanciamento daqueles locais ou maior isolamento em áreas residenciais. Fonte dos dados básicos: COVID-19 Community Mobility Reports, Google.}
    \label{fig:variacao-america-sul}
\end{figure}

A Figura \ref{fig:variacao-america-sul}  apresenta a  variação média da mobilidade no período de 15 de fevereiro a 16 de maio nos países selecionados da América do Sul. Considerando os locais que pressupõem a saída dos indivíduos de suas residências, de um modo geral, para todos os países houve uma menor variação da mobilidade em relação à linha de base para o local que inclui supermercados e farmácias. Para essa categoria da Google predominaram cores mais escuras, indicando menor distanciamento dos indivíduos para essas localidades. O mesmo ocorreu no caso dos locais de trabalho, porém em menor escala.  

Países como Brasil, Uruguai e Paraguai se destacaram com menor variação média da mobilidade em praticamente todas as categorias analisadas, e no que se refere à mobilidade em residências, essa parece ter sido uma dificuldade enfrentada por todos os países em alguma medida. 

Quando se observa a variação média da mobilidade interna aos países selecionados (Argentina, Brasil e Colômbia) verifica-se importantes tendências. Mesmo países como Argentina e Colômbia que, conforme indicado na Figura \ref{fig:america-sul-lockdown}, se situaram entre as nações com menor nível agregado de mobilidade, a redução da circulação de pessoas não ocorreu de maneira uniforme em seus territórios e entre categorias. 

\begin{figure}[H]
    \centering
    \includegraphics[width=0.9\textwidth]{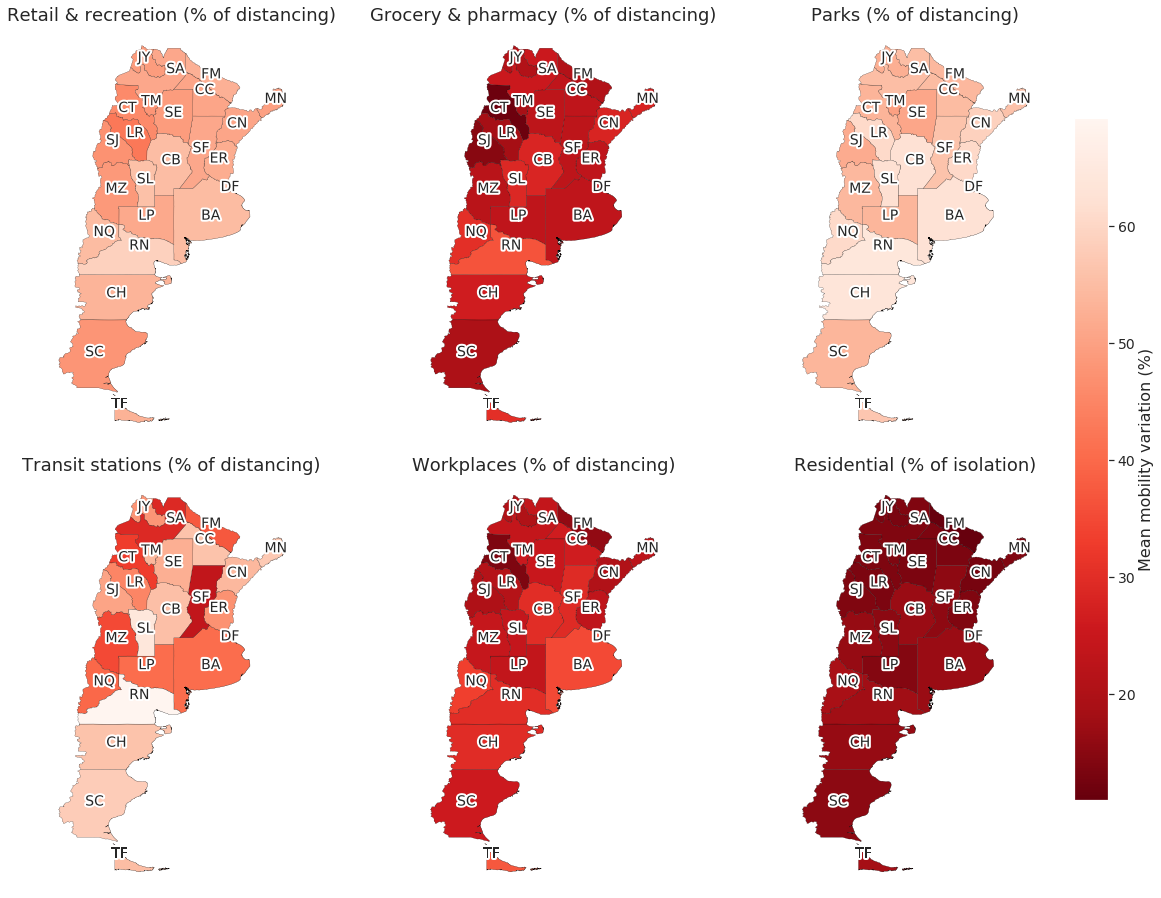}
    \caption{Variação média da mobilidade em relação à linha de base (3 de janeiro a 6 de fevereiro de 2020) para a Argentina durante o período de 15 de fevereiro a 16 de maio de 2020. Cores mais claras indicam maior distanciamento daqueles locais ou maior isolamento em áreas residenciais. Fonte dos dados básicos: COVID-19 Community Mobility Reports, Google.}
    \label{fig:variacao-argentina}
\end{figure}

No caso da Argentina (Figura \ref{fig:variacao-argentina}), a menor circulação de pessoas no período considerado pode ter sido alcançada pela redução da mobilidade em locais como compras e recreação, parques e estações de trânsito, em que foi predominante cores mais claras em seus respectivos mapas. 

\begin{figure}[H]
    \centering
    \includegraphics[width=0.9\textwidth]{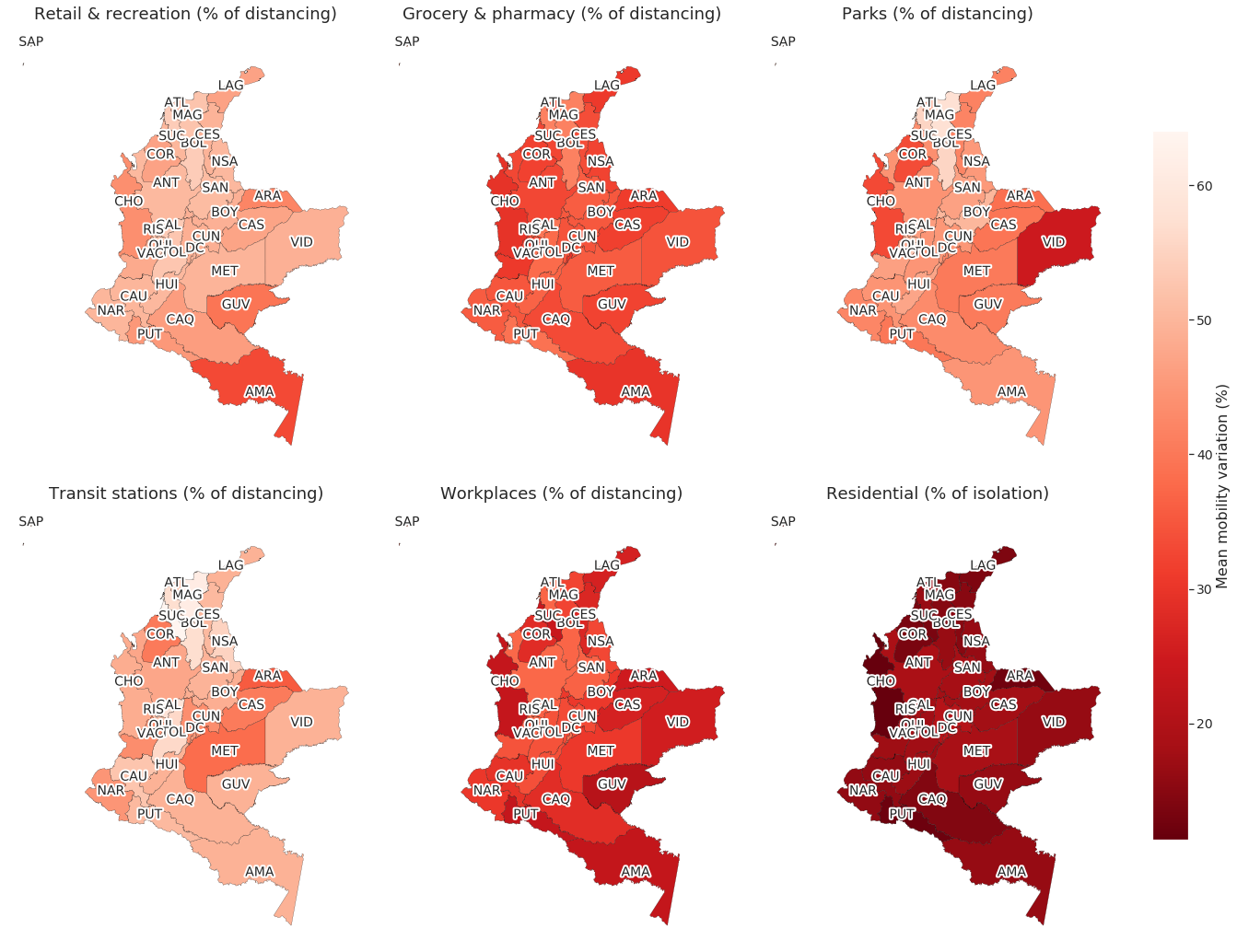}
    \caption{Variação média da mobilidade em relação à linha de base ( 3 de janeiro a 6 de fevereiro de 2020) para a Colômbia durante o período de 15 de fevereiro a 16 de maio de 2020. Cores mais claras indicam maior distanciamento daqueles locais ou maior isolamento em áreas residenciais. Fonte dos dados básicos: COVID-19 Community Mobility Reports, Google.}
    \label{fig:variacao-colombia}
\end{figure}

No caso da Colômbia (Figura \ref{fig:variacao-colombia}), houve maior variabilidade interna quanto à mobilidade nas seis categorias considerados pela Google, e com mapas descrevendo tendências menos uniformes do que no caso da Argentina. Os departamentos colombianos apresentaram cores mais claras para a categoria que expressa a mobilidade em locais de estações de trânsito, e a redução da circulação de pessoas nessas áreas pode ter sido importante para que esse país se situasse entre aqueles com menor nível de mobilidade durante boa parte do período considerado. 

\begin{figure}[H]
    \centering
    \includegraphics[width=0.9\textwidth]{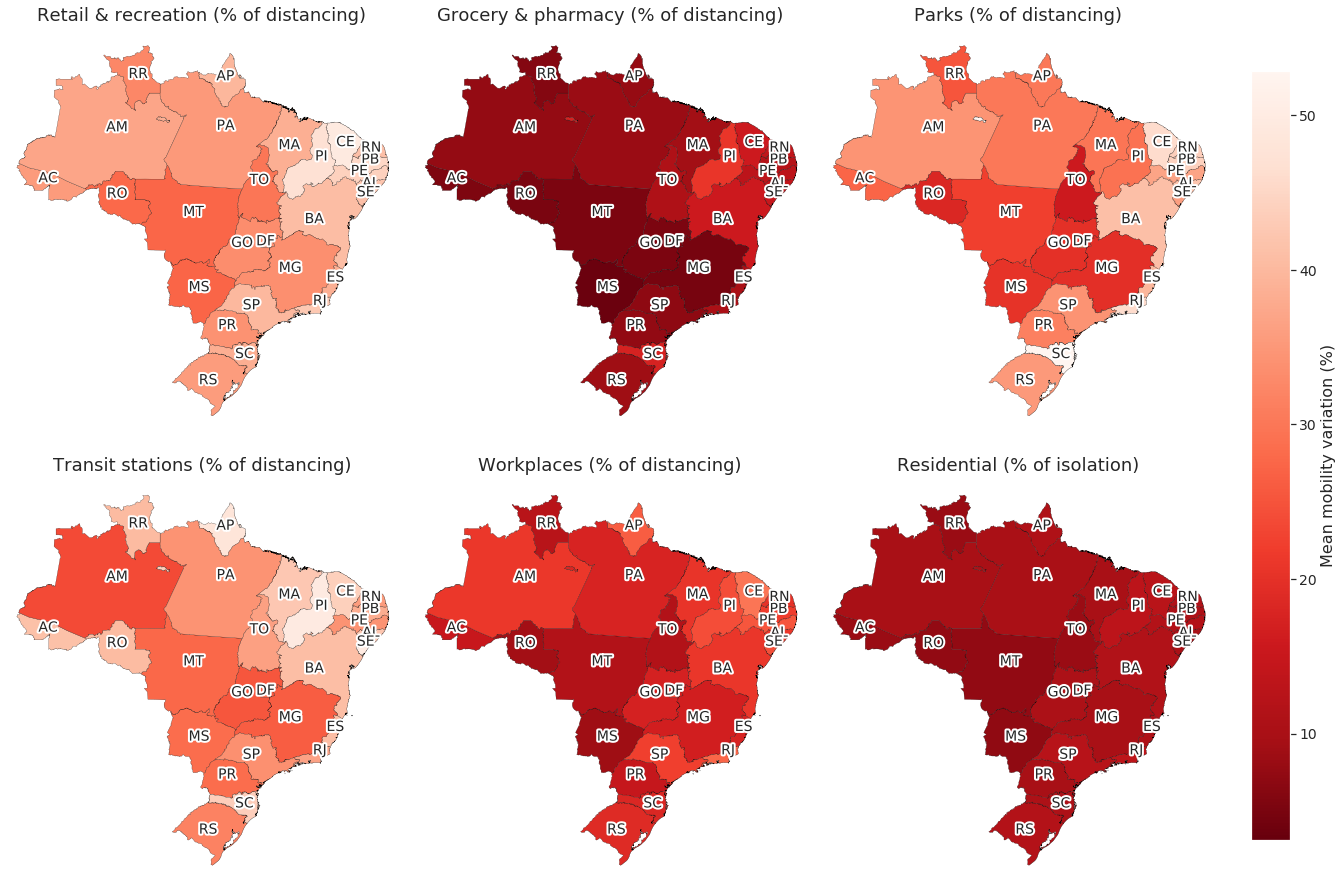}
    \caption{Variação média da mobilidade em relação à linha de base (3 de janeiro a 6 de fevereiro de 2020) para o Brasil durante o período de 15 de fevereiro a 16 de maio de 2020. Cores mais claras indicam maior distanciamento daqueles locais ou maior isolamento em áreas residenciais. Fonte dos dados básicos: COVID-19 Community Mobility Reports, Google.}
    \label{fig:variacao-brasil}
\end{figure}

No caso do Brasil, a Figura \ref{fig:variacao-brasil} mostra predomínio de cores mais escuras em locais de trabalho, supermercados e farmácias, e residências. Esse país adotou medidas pontuais de \textit{lockdown} e somente no terceiro mês pós pandemia, e conforme ilustrado pela Figura \ref{fig:america-sul-lockdown}, apresentou o maior nível de circulação de pessoas durante praticamente todo o período analisado. O quesito relacionado às compras e recreação, que incluem o funcionamento de \textit{shoppings centers} foi o que apresentou maior predomínio de cores claras no mapa, indicando maior variação média em relação ao período de linha de base. Essa tendência é compatível com as diretrizes de boa parte dos decretos estaduais do Brasil que regulamentou o encerramento de atividades nesses espaços, sobretudo nos dois primeiros meses da pandemia. Já em locais em que não houve maiores restrições quanto ao acesso da população, como farmácias e supermercados, cores mais escuras no mapa predominaram, indicando menor variabilidade na circulação de pessoas em relação ao período pré-pandemia, e praticamente de norte a sul do país.

\subsection{Análise de autocorrelação espacial descritiva}
\subsubsection{Índice de Moran Global}
\begin{figure}[ht]
    \centering
    \includegraphics[width=0.9\textwidth]{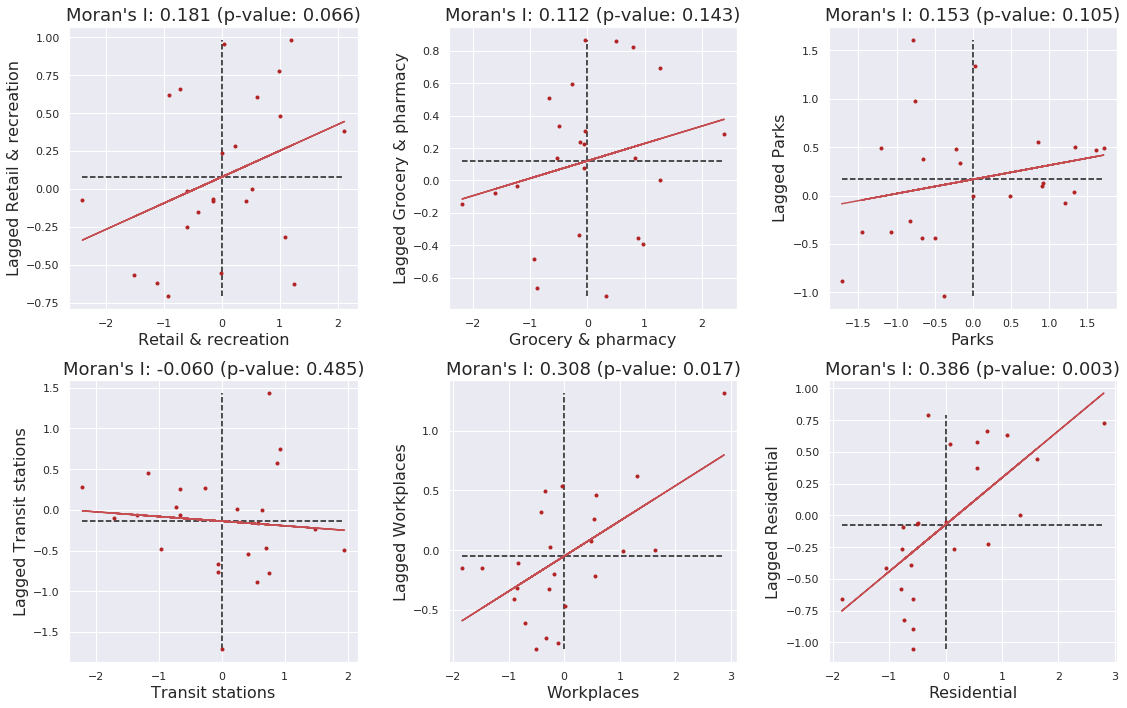}
    \caption{Diagramas de Espalhamento de Moran para as categorias da Google: Argentina. Fonte dos dados básicos: COVID-19 Community Mobility Reports, Google.}
    \label{fig:moran-global-argentina}
\end{figure}

\begin{figure}[H]
    \centering
    \includegraphics[width=0.9\textwidth]{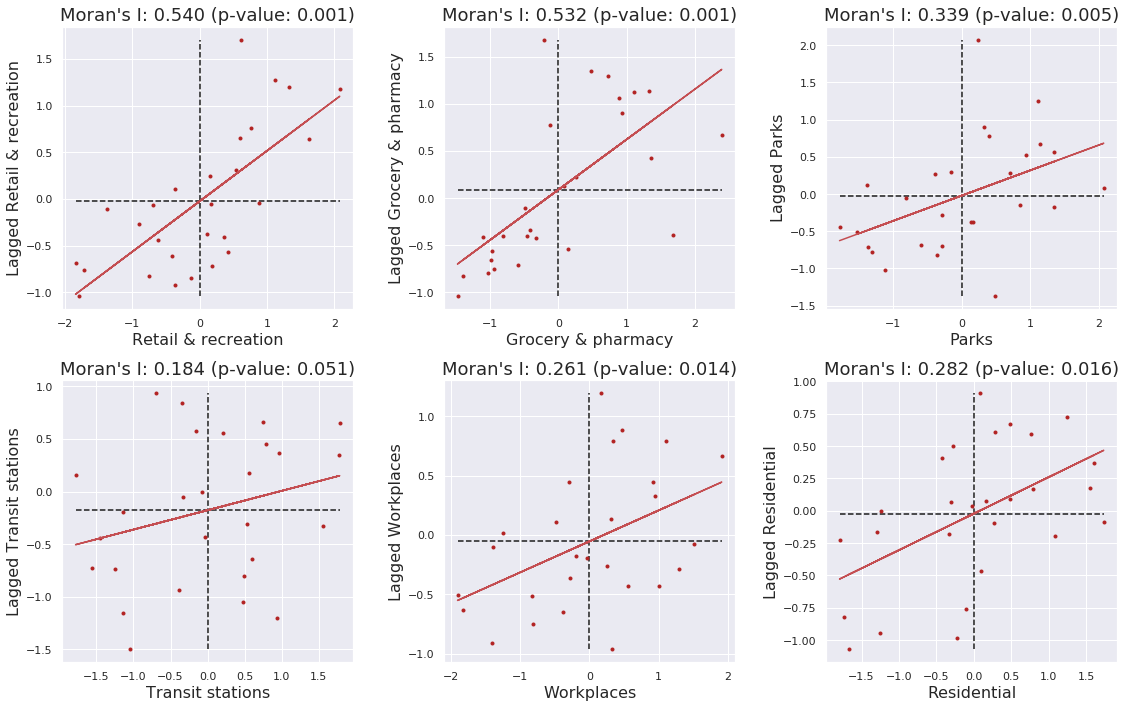}
    \caption{Diagramas de Espalhamento de Moran para as categorias da Google: Brasil. Fonte dos dados básicos: COVID-19 Community Mobility Reports, Google.}
    \label{fig:moran-global-brasil}
\end{figure}

As Figuras de \ref{fig:moran-global-argentina} a \ref{fig:moran-global-brasil} ilustram os Diagramas de Espalhamento de Moran para Argentina, Colômbia e Brasil, respectivamente, a fim de verificar se há padrões de aglomeração espacial para a mobilidade nos três países selecionados.

Iniciando pela Argentina, os Índices de Moran Global estatisticamente significativos ao nível de 95\% de confiabilidade foram os locais de trabalho e residência. Portanto, ao menos para essas duas categorias, não se pode dizer que a mobilidade da população de 15 de fevereiro a 16 de maio ocorreu de maneira aleatória (Figura \ref{fig:moran-global-argentina}). 

Ainda com relação à Figura \ref{fig:moran-global-argentina}, no que se refere ao Índice de Moran Global para a categoria local de trabalho e residência, pode-se dizer que a autocorrelação espacial foi positiva, ou seja, uma localização possui vizinhos com valores semelhantes. No caso do valor dessa estatística para a categoria que expressa a mobilidade dos indivíduos em locais residenciais, a força da associação foi ligeiramente maior em comparação ao valor desse índice para a categoria locais de trabalho.  

\begin{figure}[H]
    \centering
    \includegraphics[width=0.9\textwidth]{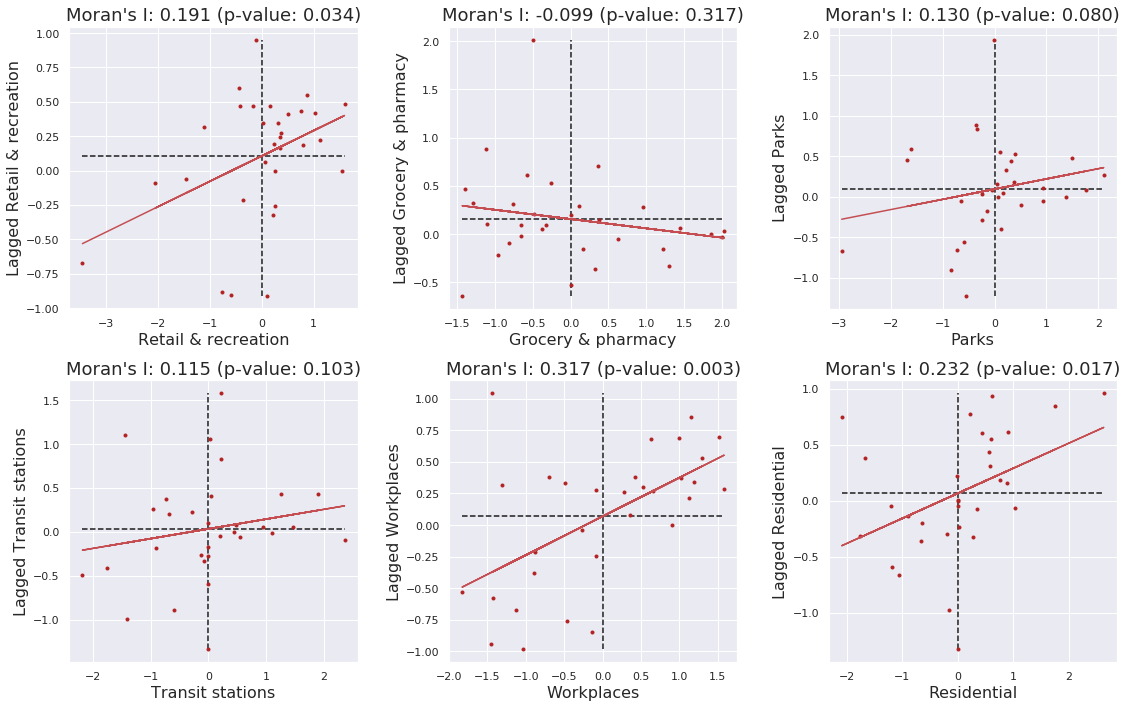}
    \caption{Diagramas de Espalhamento de Moran para as categorias da Google: Colômbia. Fonte dos dados básicos: COVID-19 Community Mobility Reports, Google.}
    \label{fig:moran-global-colombia}
\end{figure}

Já para a Colômbia, além de locais de trabalho e residência houve também a formação de agrupamentos para a mobilidade em locais de compras e recreação (Índices de Moran Global estatisticamente significativos). Assim como observado no caso da Argentina, o indicador global apontou para uma autocorrelação espacial positiva e mais forte para a mobilidade em residências (Figura \ref{fig:moran-global-colombia}).

No caso do Brasil (Figura \ref{fig:moran-global-brasil}), o Índice de Moran Global alcançou significância estatística em todas as categorias de locais da Google, além de apontar para uma autocorrelação espacial positiva e mais forte do que foi verificado para a Argentina e a Colômbia. 

Em conjunto, os resultados encontrados para os três países corroboraram a hipótese de que há a formação de padrões regionais de mobilidade nos países selecionados da América do Sul (Argentina, Brasil e Colômbia). Ela foi confirmada de maneira mais categórica para o Brasil, cujos resultados do Índice de Moran Global indicaram para todos os locais que a mobilidade da população no período considerado seguiu um padrão espacial, ou seja, não ocorreu por um mero acaso.

\subsubsection{Índice de Moran Local}
Realizada a análise do indicador global de autocorrelação espacial, esse artigo apresenta também a análise do índice de Moran local que aponta as províncias/departamentos ou estados da Argentina, Colômbia e Brasil que apresentaram correlação local significativamente distinto das demais. 

\begin{figure}[ht]
    \centering
    \includegraphics[width=0.9\textwidth]{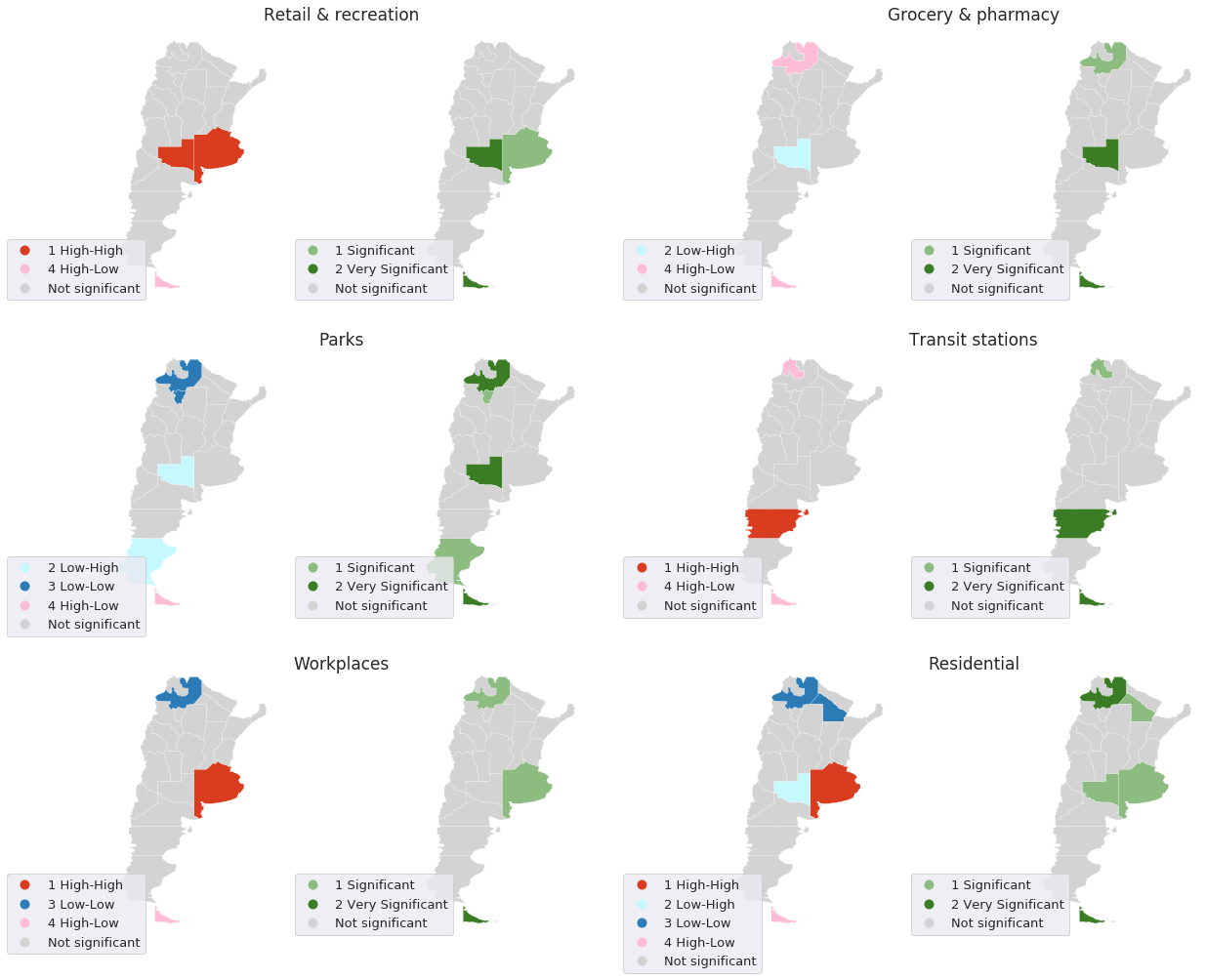}
    \caption{Mapas de \textit{cluster} e de significância LISA para as categorias da Google: Argentina. Fonte dos dados básicos: COVID-19 Community Mobility Reports, Google.}
    \label{fig:moran-local-argentina}
\end{figure}

Conforme ilustrado pela Figura \ref{fig:moran-local-argentina}, no caso da Argentina, se formaram agrupamentos pequenos, envolvendo no máximo duas províncias. Em pelo menos quatro das seis categorias de locais da Google, se formaram agrupamentos do tipo \textit{high-high}, ou seja, localidades de alta variação da mobilidade no período de 15 de fevereiro a 16 de maio e em relação ao período de linha de base (cores vermelhas no mapa). Em três desses agrupamentos (compras e recreação, locais de trabalho e residência) a província de Buenos Aires que representa a mais populosa e com muitos casos da COVID-19
~\footnote{\url{https://g1.globo.com/bemestar/coronavirus/noticia/2020/05/24/argentina-prorroga-lockdown-ate-7-de-junho-e-endurece-medidas-em-buenos-aires.ghtml}}, esteve incluída.

Também é possível identificar na Figura \ref{fig:moran-local-argentina}, agrupamentos do tipo \textit{low-low}, ou seja, aglomerações de províncias com baixa variação da mobilidade em relação ao período pré-pandemia na América do Sul. Isso ocorreu para os locais de parques, locais de trabalho e residência, sendo que em todas a província de Salta localizada no Noroeste da Argentina esteve presente. 

Verificou-se também a formação de \textit{outliers} para a Argentina: locais com baixa variação da mobilidade cercados por vizinhos com alta variação (\textit{low-high}), como o \textit{outlier} formado por La Pampa para o local de residência e que se situou entre um agrupamento do tipo \textit{high-high}, no caso a província de Buenos Aires; e locais com alta variação da mobilidade cercados por vizinhos com baixa variação (\textit{high-low}), como o arquipélago de Tierra del Fuego que por suas características insulares formou um agrupamento dessa natureza para as seis categorias de mobilidade da Google (Figura \ref{fig:moran-local-argentina}).

\begin{figure}[H]
    \centering
    \includegraphics[width=0.9\textwidth]{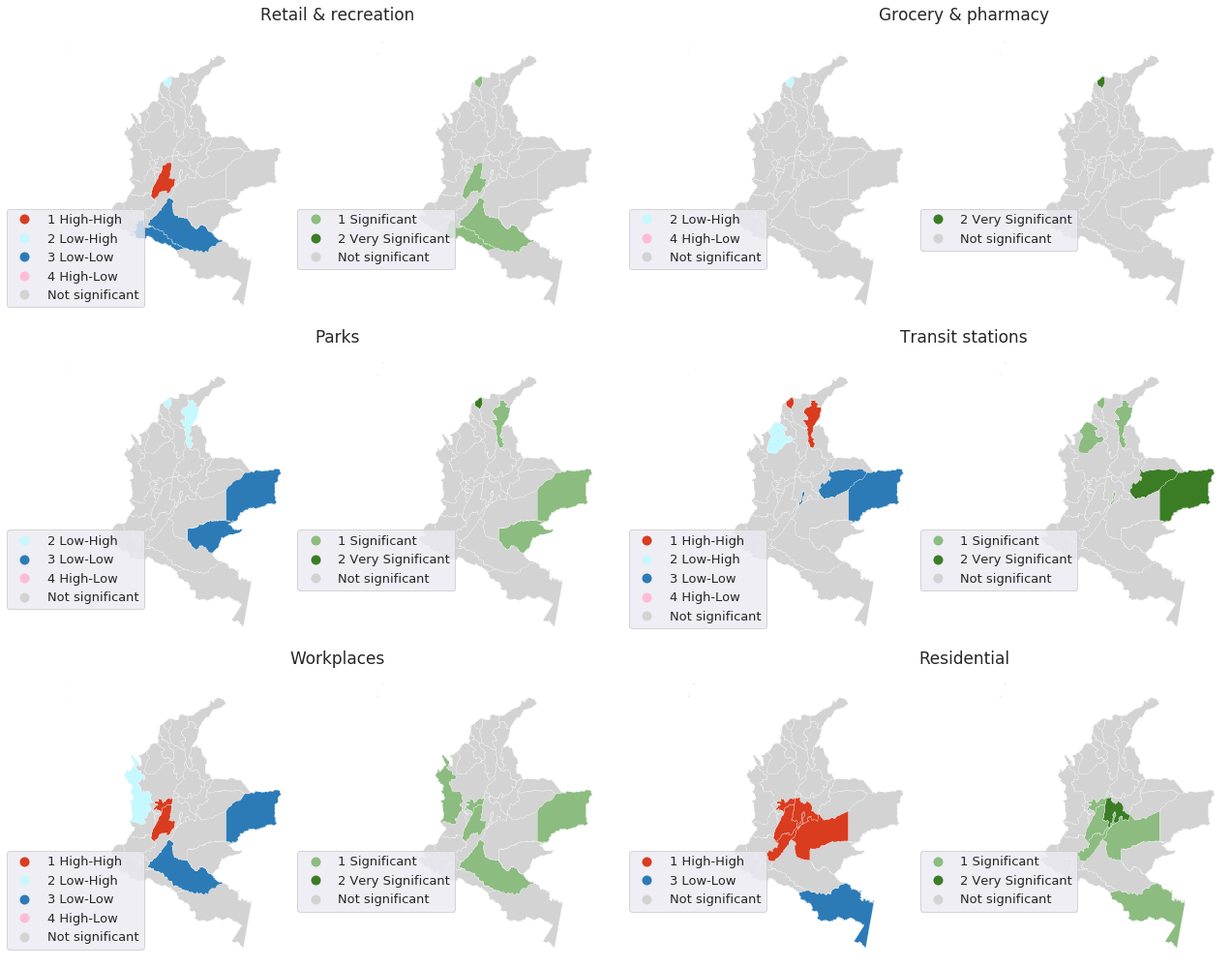}
    \caption{Mapas de \textit{cluster} e de significância LISA para as categorias da Google: Colômbia. Fonte dos dados básicos: COVID-19 Community Mobility Reports, Google.}
    \label{fig:moran-local-colombia}
\end{figure}

No caso da Colômbia, a Figura \ref{fig:moran-local-colombia} também mostra a formação de pequenos agrupamentos, tendo sido o maior deles observado para a categoria residencial que envolveu quatro departamentos e formou um agrupamento de alta variação na mobilidade (\textit{high-high}): Meta, Huila, Cundinamarca e Tolima. Em Bogotá, capital da Colômbia, se concentram a maior  parte dos óbitos por COVID-19~\footnote{\url{https://noticias.uol.com.br/ultimas-noticias/efe/2020/05/06/colombia-bate-recorde-diario-de-casos-de-coronavirus.htm}}. Agrupamentos do tipo \textit{low-low} foram os mais frequentes (cinco das seis categorias de espaços físicos da Google) assim como os \textit{outliers low-high}, ou seja, locais com baixa variação da mobilidade cercados por vizinhos com alta variação da circulação de pessoas.

\begin{figure}[ht]
    \centering
    \includegraphics[width=0.9\textwidth]{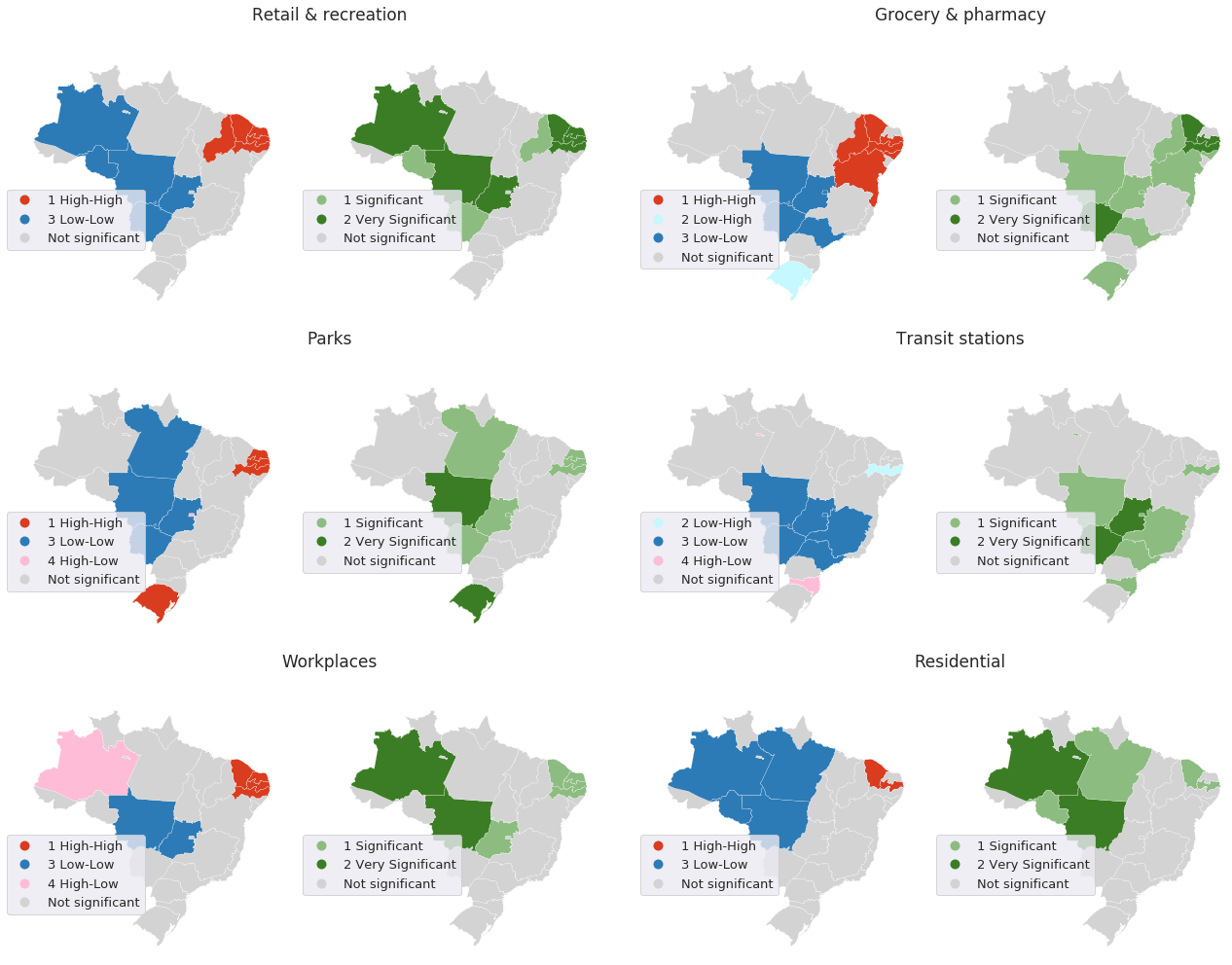}
    \caption{Mapas de \textit{cluster} e de significância LISA para as categorias da Google: Brasil. Fonte dos dados básicos: COVID-19 Community Mobility Reports, Google.}
    \label{fig:moran-local-brasil}
\end{figure}

Para o Brasil, observa-se a formação de agrupamentos envolvendo um número maior de divisões territoriais do que o observado para Argentina e Colômbia. Tendo em vista que para o Brasil a autocorrelação espacial indicada pelo Índice de Moran Global foi mais forte do que para os outros dois países analisados, a formação de agrupamentos mais bem definidos foi um resultado esperado (Figura \ref{fig:moran-local-brasil}).

Predominaram agrupamentos \textit{low-low} (baixa variação da mobilidade em relação à linha de base) que envolveram na maior parte deles estados da Região Centro-Oeste, da Região Norte e o estado de São Paulo, localizado na Região Sudeste do país. Para a categoria compras e recreação formou-se um agrupamento \textit{low-low} com cinco estados, sendo todos da Região Centro Oeste e dois estados da Região Norte. Essas regiões representam, na média, áreas de de cobertura da Floresta Amazônica e de baixa densidade populacional, como o estado do Amazonas localizado ao norte do país (Figura \ref{fig:moran-local-brasil}). 

No caso do Brasil, chama a atenção também o fato dos agrupamentos \textit{high-high} (alta variação da mobilidade em relação à linha de base) envolverem em sua maioria (cinco de seis categorias) estados da Região Nordeste do país. Até maio, alguns estados dessa localidade já se encontravam duramente afetados pela pandemia por COVID-19, e alguns de seus municípios mais importantes se preparavam para entrar em \textit{lockdown} (Figura \ref{fig:moran-local-brasil}).

Com relação aos \textit{outliers}, no Brasil eles foram menos frequentes do que na Argentina e na Colômbia, o que é coerente com o tipo de distanciamento social mais rígido adotado por essas nações em relação ao que foi adotado no Brasil: sob um regime de \textit{lockdown} comportamentos excepcionais da mobilidade em relação à maioria (\textit{outliers}) são mais esperados do que em regimes mais flexíveis de medidas de distanciamento. Também para o caso brasileiros, foram identificados agrupamentos \textit{high-low} para locais de trabalho (estado do Amazonas) e estações de trânsito (Santa Catarina); e agrupamentos \textit{low-high} para supermercados e farmácias (Rio Grande do Sul) e estações de trânsito (Pernambuco).

\section{Discussão}
Em um contexto em que não se dispõe de recursos farmacêuticos para conter o avanço do coronavírus em todo o mundo, o distanciamento social ocupa posição de centralidade nas ações de mitigação da COVID-19. Utilizando dados de geolocalização de celulares do Community Mobility Reports da Google, foi possível identificar as variações na circulação de pessoas em países selecionados da América do Sul para diferentes modalidades de distanciamento social adotadas durante a pandemia por COVID-19.  

Considerado um período de cerca de três meses (15 de fevereiro e 16 de maio) e utilizando análises descritivas para descrever as tendências de mobilidade de dez países da região por meio de um indicador sintético e, também, para identificar a existência de autocorrelação espacial nas tendências de circulação para três nações que adotaram medidas de distanciamento social distintas, sobretudo no início da pandemia, foi possível testar duas hipóteses: (a) em países com decreto de \textit{lockdown} há maior sucesso na redução da mobilidade da população e (b) países selecionados (Argentina, Brasil e Colômbia) formam padrões regionais de mobilidade. 

Sobre a primeira hipótese, a tendência descrita pelo indicador de circulação dividiu os países analisados em dois grupos, tendo sido menor a circulação de pessoas para a maior parte dos países que adotaram o \textit{lockdown} (menor circulação). Nesse caso a exceção foi o Paraguai, que apesar de ter adotado essa medida extrema de distanciamento social em 20 de março, se situou na maior parte do tempo pós-pandemia no grupo dos países com maior nível de mobilidade. Nessa nação, até dia 06 de junho apresentava-se um dos menores níveis de infecção pelo novo coronavírus da América do Sul, com pouco mais de mil casos confirmados e apenas 11 óbitos \cite{WHO-136}. As características peculiares de população e extensão territorial do Paraguai frente a de seus pares sul americanos, pode ter compensando a tendência oscilante na circulação de pessoas observada para esse país e impedido que a doença se espalhasse com maior velocidade. Portato, a excepcionalidade da tendência seguida pelo Paraguai permitiu confirmar apenas de maneira parcial a hipótese de que em países com decreto de \textit{lockdown} há maior sucesso em reduzir a mobilidade da população. 

Os resultados obtidos por meio da análise descritiva de autocorrelação espacial (Índice de Moran Global e Local) mostraram que há dependência no espaço quanto à mobilidade da população no período considerado para os três países, confirmando, portanto, a nossa segunda hipótese. Sobretudo no caso do Brasil, país com o maior nível de mobilidade após a tomada das primeiras medidas de distanciamento social na região como um todo, a dependência espacial foi mais forte do que na Argentina e na Colômbia, países que estiveram na maior parte do tempo sob um regime estrito de \textit{lockdown}. Este resultado é condizente com as medidas de distanciamento social adotadas por esses países, ou seja, em um cenário menos rígido de determinações nacionais restringindo a circulação de pessoas como no caso do Brasil, as unidades político-administrativas apresentam maior flexibilidade para lidar com seus decretos, o que se refletiu na formação de grandes aglomerações regionais nas tendências de circulação durante a pandemia. Em um cenário mais rígido, padrões regionais de mobilidade parecem importar menos ou sequer são formados. Por exemplo, no caso da Argentina e da Colômbia, a formação de \textit{outliers} foi mais comum do que no caso brasileiro, assim como a formação de agrupamentos pequenos, ou seja, envolvendo poucos departamentos/províncias. Cumpre destacar que nesse artigo não foram considerados detalhes quanto às medidas adotadas pelos países para avaliar seus efeitos sobre a circulação de pessoas, apenas a natureza dos decretos do poder executivo de cada país, ou seja, se houve ou não o \textit{lockdown}. 

Importante considerar também que características culturais, sociais, econômicas e políticas entre os países podem ter contribuído para os resultados encontrados e devem ser levados em consideração para se analisar a adesão às medidas de distanciamento social durante a pandemia. O Brasil que representa um caso à parte na América do Sul por suas dimensões continentais e grande diversidade sociocultural, também se destaca frente aos demais países quanto ao avanço da COVID-19. Conforme demonstrado no presente estudo, a alta circulação de pessoas e a formação de grandes agrupamentos do tipo \textit{low-low} (baixa variação na mobilidade em relação à linha de base) indicam que decretos menos rígidos podem não ser por si mesmos eficientes para conter a circulação de pessoas. Em 06 de junho, o Brasil era o líder quanto ao total de novos casos  na região, e o número de óbitos naquele país superava 30 mil \cite{WHO-136}.  Conforme já mencionado, somente no terceiro mês da pandemia alguns estados brasileiros adotaram períodos curtos de \textit{lockdown}, sobretudo em cidades de maior porte, como Pernambuco e Fortaleza, ambos localizados na Região Nordeste do país e que enfrenta um rápido avanço da COVID-19.  

Importante destacar também que em ambos os países, os agrupamentos que indicam maior variação na mobilidade envolveram estados/províncias com nível elevado de infecção pelo novo coronavírus, como a província de Buenos Aires, a capital Bogotá, e os estados da Região Nordeste do Brasil. Pode ser que em locais duramente atingidos a população siga com maior rigor os decretos que inibem a mobilidade, e independentemente do tipo de decreto de distanciamento em vigor, se mais ou menos rígido. 

\section{Limitações do estudo e considerações finais}
Com relação aos dados e a metodologia utilizada, apesar das informações analisadas terem propiciado o alcance de resultados consistentes, é importante considerar que os dados da Google apresentam limitações que também devem ser observadas. Por exemplo, se desconhece para cada unidade geográfica a cobertura da população de usuários que ativam o histórico de localização de seus dispositivos móveis e que permitem que a empresa capture dados de mobilidade desses indivíduos. A esse respeito, a informação disponível pela Google em sua documentação metodológica, é que em localidades e categorias que não alcançaram níveis suficientes de dados estatisticamente significativos não foram incluídos nas séries temporais disponibilizadas pela empresa~\footnote{\url{https://www.google.com/covid19/mobility/}}. Ainda com relação às possíveis limitações das categorias de espaços físicos da Google, locais como residências e de trabalho são difusas, o que pode conduzir a resultados imprecisos quanto à população que realmente praticou o distanciamento social permanecendo em suas casas ou que de fato desempenharam atividades laborais durante a pandemia. 

Quanto à metodologia empregada, é possível que a escolha por análises espaciais ao nível de províncias/departamentos ou estados possa não contemplar a variabilidade interna dos países quanto à mobilidade, sendo importante avançar em dados que permitam o acompanhamento do distanciamento social em menores unidades de análise. Além do emprego da mesma metodologia para outros países da região, e para os quais se disponha de informações intranacionais para verificar se o mesmo padrão encontrado para os três países analisados se mantém. Sugere-se também que trabalhos futuros explorem períodos de tempo mais curtos e que acompanhe as mudanças nos decretos nacionais, a fim de se verificar se os padrões de autocorrelação espacial se alteram com as mudança de diretrizes dos governos quanto ao distanciamento social.

Todavia, considera-se que esse trabalho contribuiu para o entendimento de que ao menos para América do Sul, medidas de \textit{lockdown} ou de distanciamento social menos rígidas são necessárias, porém, não são suficientes para a redução da circulação de pessoas durante a pandemia por COVID-19.

\vspace{6pt}

\acknowledgments{O presente trabalho foi realizado com apoio da Coordenação de Aperfeiçoamento de Pessoal de Nível Superior - Brasil (CAPES) - Código de Financiamento 001.}

\conflictsofinterest{The authors declare no conflict of interest.} 

\reftitle{References}


\externalbibliography{yes}
\bibliography{ref.bib}



\end{document}